\documentstyle[preprint,aps]{revtex}

\font\tenrm=cmr10

\font\bmfont=cmbxti10 scaled \magstep2
\def\bm#1{\mbox{\bmfont #1}}

\if@twoside
\else \oddsidemargin 00pt \evensidemargin 00pt
\fi
\expandafter\ifx\csname mathrm\endcsname\relax
\def\mathrm#1{{\rm #1}}\fi

\makeatother

\def\limit#1#2{\ \rlap{\raise 2pt \hbox{$\ \longrightarrow$}}
                    {\lower 2pt \hbox{$\scriptstyle#1
                      \rightarrow#2$}}\ }
\def\to{\rightarrow}

\newcommand{\beq}{\begin{equation}}
\newcommand{\eeq}{\end{equation}}
\newcommand{\beqa}{\begin{eqnarray}}
\newcommand{\eeqa}{\end{eqnarray}}
\newcommand{\ben}{\begin{enumerate}}
\newcommand{\een}{\end{enumerate}}
\newcommand{\bei}{\begin{itemize}}
\newcommand{\eei}{\end{itemize}}
\newcommand{\bef}{\begin{figure}}
\newcommand{\eef}{\end{figure}}

\newcommand{\ind}[1]{}
\newcommand{\zpsfig}[1]{}

\newcommand{\bsg}{\mbox{$b \to s \gamma$}}
\newcommand{\bsnn}{\mbox{$b \to s \nu \bar\nu$}}

\newcommand{\bsll}{\mbox{$b \to s\, \ell^+ \ell^-$}}

\newcommand{\tcz}{\mbox{$t c Z $}}
\newcommand{\bsz}{\mbox{$b s Z $}}
\newcommand{\ul}{\mbox{${\bf U}$}}

\newcommand{\el}{\mbox{$\varepsilon_L$}}
\newcommand{\er}{\mbox{$\varepsilon_R$}}
\newcommand{\elij}{\mbox{$\varepsilon_L^{ij}$}}

\def\ea{{et al.}}
\def\ib{{\it ibid.}}

\newcommand{\Pt}{\mbox{$P_{{\scriptscriptstyle T}_3} $}}

\def\npb#1{Nucl. Phys. {\bf B #1}}
\def\plb#1{Phys. Lett. {\bf B #1}}
\def\prd#1{Phys. Rev. {\bf D #1}}
\def\prl#1{Phys. Rev. Lett. {\bf #1}}

\begin{document}

\input feynman
 \draft
{\tighten
\preprint{\vbox{\hbox{WIS-95/40/Aug-PH}
                \hbox{hep-ph/9509233}
              %  \hbox{\today}
                                }}
\title{Top-Charm flavor changing contributions \break
 to the  effective  \bm{bsZ} vertex}
\author{ Enrico Nardi\,}
\address{ \vbox{\vskip 0.3truecm}
  Department of Particle Physics \\
      Weizmann Institute of Science, Rehovot 76100, Israel }

\maketitle

\begin{abstract}%
We analyze the effects of a tree level flavor changing $tcZ$ vertex
induced by a mixing with new isosinglet $Q=2/3$ quarks, on
the effective $bsZ$ vertex. We compute  the contributions arising
from the new electroweak penguin diagrams involving one insertion of
the $tcZ$ vertex. We show that a generalized GIM mechanism ensures
the cancellation  of the mass independent terms as well as of the
new divergences. Unexpectedly, the presence of a $tcZ$ coupling cannot
enhance the rates for the $Z$ mediated flavor changing decays \bsll\
and \bsnn, implying that these processes cannot be used to set limits
on the $tcZ$ coupling. The additional effects of the heavy isosinglets
are compared with the well studied effects of new isodoublets appearing
in multi-generational models.
\end{abstract}
\vfill
\noindent
--------------------------------------------\phantom{-} {\hfill\break}
Electronic mail:\quad ftnardi@wicc.weizmann.ac.il
}
%%%%%%%%%%%%%%%%%%%%%%%%%%%%%%%%%%%%%%%%%%%%
\newpage
%%%%%%%%%%%%%%%%%%%%%%%%%%%%%%%%%%%%%%%%%%%%%%

\baselineskip 18pt

% \end{document}

In the Standard Model (SM)  Flavor Changing Neutral
Current (FCNC) processes are strongly suppressed due to the GIM
 mechanism. The experimental confirmation of this suppression can
 be regarded as one of the successes of the theory, and at the same
 time it represents a challenging constraint for most new physics
 scenarios which often predict new sources of FCNC.
In several new physics models, the new sources of  FCNC are
 related to ratios between the
masses of the fermions involved in the FC transitions and some new
 mass scale, of the order of the electroweak breaking scale or
 larger. This is the case for example in models where FCNC arise from
 a mixing between the light fermions and new heavy states with
 non-standard $SU(2)_L$ assignments  \cite{ll-nrt,FCmix,burgess,buchi},
in multi Higgs doublets models without natural flavor conservation
 \cite{topFCNC,Hall-Weinberg} or in models which try to explain the
 fermion mass hierarchy by means of horizontal symmetries \cite{QSA}.
Due to the smallness of the  fermion masses,
in all these cases the new FCNC effects  are naturally suppressed.

However, if this is the underlying mechanism responsible for the
observed suppression, then the absence of FCNC at low energies does
 not imply the same suppression at large mass scales.
In particular,  due to the large value of $m_t$
such a suppression might not be effective for FC transitions
 involving the top quark \cite{topFCNC,Hall-Weinberg}.
Recently some attention has been paid to  study
this kind of $t\,c$ FC transitions,   both from the point of view
 of model building \cite{topFCNC,Hall-Weinberg}
 as well as from the point of view of the possible phenomenological
 consequences \cite{MHDM-FCNC,laura,peccei}.

In this letter we investigate the consequences of a tree level
  $tcZ$ FCNC vertex  arising from a mixing
 between  the  known $u$-type  quarks and {\it new} $Q=2/3$
isosinglet heavy states on the effective $bsZ$ vertex.
In spite of the loop suppression, there are good reasons for
 carrying out an analysis of these  effects.
Under general assumptions,  the strength  of the $u_i\,u_j$ FCNC
 coupling to the $Z$ boson
induced by a mixing  is  expected to be
of  order $ \sim {m_i m_j}/M^2$,
where $M$ is the mass-scale of the new states \cite{parada}.
In this case we would expect that
the contribution to the \bsz\ vertex induced by penguin diagrams with
 one insertion of the  \tcz\ vertex, could  be
even larger  than a  \bsz\ tree
level coupling arising from a similar mechanism, that is from
a mixing in the $d$-quark sector with new $Q=-1/3$ isosinglets.
Namely, we would expect  the ratio between the loop induced and the tree level
 $bsZ$ vertices to be
\beq
{\Gamma_{\bsz}^{\rm penguin}\over \Gamma_{\bsz}^{\rm tree}} \sim
{V^*_{tb}V_{cs}\over (4 \pi)^2} {m_t m_c\over m_b m_s}
\sim 2.5
\eeq
implying that  the sensitivity of the \bsz\ effective vertex
to FC mixing effects could be mainly determined
by the presence of a \tcz\ coupling.

Apart from inducing FC couplings, a mixing with new isosinglets
 quarks leads also to the non-unitarity of the $3\times 3$
Cabibbo-Kobayashi-Maskawa (CKM) matrix  in the same way as  the
 presence of a fourth generation does.
Therefore,  the  effects induced by  one additional $Q=2/3$
 isosinglet in processes like   $B_d^0-\bar B_d^0$ mixing and
 $b\to s \gamma$, which are essentially related to the non-unitarity
of the CKM matrix,  are the same as from a fourth generation,
and the same constraints apply in both cases.
These effects  have been recently  analyzed in \cite{barger}.
The effects of a fourth generation on $b\to s \gamma$
were previously studied in \cite{soni}, and  an
analysis of the constraints on a fourth generation implied by the
experimental measurement of this decay \cite{CLEO-bsg-in},
which  applies for the isosinglet case as well,
has been presented in \cite{hewett}.
The result is that after imposing  the limits
 on  the CKM matrix deviations from unitarity,
 the presence  of a new $Q=2/3$ isodoublet
(or  isosinglet)  quark  in the mass range 200-400 GeV
is still consistent with the  measured rate for
 $b\to s \gamma$ \cite{CLEO-bsg-in}.
Limits on the masses of new fermions in
additional generations  have been derived also from analyses of
the precise electroweak data \cite{burgess,fourth-fam}, by
constraining the contribution to  the electroweak radiative
 corrections  from  the additional doublets.
However,  the isosinglets are coupled to the $SU(2)$ gauge bosons
 only through small mixings with the standards quarks, and therefore
 these additional constraints do not apply to the isosinglet case.

With regards to the rare FC decays,
the isosinglet case differs from a four generation SM due to the
possible presence of   FCNC couplings that, as we have stressed,
can be particularly large for the $t$ quark.
Since $Z$ exchange does not contribute either to  $B^0-\bar B^0$
 mixing or to $b\to q \gamma$,  both these processes are not
 sensitive to these couplings.
However, both the FC decays  \bsll\  and \bsnn, which are strongly
 suppressed in the SM, are  sensitive to $Z$ exchange. In this case
 the presence  a  $tcZ$ vertex gives rise to  new electroweak
penguin diagrams which, being proportional to  $V^*_{tb}V_{cs} $,
are not expected to be suppressed by small CKM mixings.
The aim of our analysis is to see whether
it is possible to bound a FC $tcZ$ coupling induced by a mixing with
 isosinglets  by searching for these rare decay modes.
Our main results are the following:

\begin{itemize}
\item{
The size of the contribution  to the effective $bsZ$ vertex
of the new penguin diagrams induced by a $tcZ$ vertex
is bounded to be smaller than the SM result, and interferes
 destructively with it. Hence the rate for the  FC decays is lowered
 by this effect, and the  experimental upper limits on \bsll\
 \cite{UA1} and on \bsnn\ \cite{yze} do not imply any constraint on
a mixing induced  $tcZ$ vertex.
}
\item{
The additional effects due to the new  diagrams involving loops of
 the heavy singlets can give an enhancement to the decay rates.
 However, in the limit of large masses,  the
isosinglet nature of the new states yields only small
 logarithmic corrections to the known result for
new isodoublets with the same masses \cite{inami-lim}.
Therefore any signature of the presence of  additional $Q=2/3$
 isosinglets in future measurements of the  \bsll\  and \bsnn\ decay
 modes, if detected,  will not be easily distinguished from the
contributions of a heavy fourth generation.
}
\end{itemize}

We will first briefly present the general formalism for describing the
 effects of $u$-quark mixing with  new isosinglets.
Then we will generalize the SM   computation  of the effective \bsz\
vertex \cite{inami-lim} to the  case when fermion mixing induces
tree level FCNC couplings  in  the $u$-quark sector. We will show that
in this case a generalized GIM mechanism
ensures the finiteness of the result with no need of introducing
a cut-off by hand. We note that in general a cut-off is still needed
in other cases when a $tcZ$  coupling  does not arise at tree
 level, but is generated as an effective vertex \cite{peccei}.
Finally, we will discuss the phenomenological implications of our
 results. Our analysis  complements, and sometimes parallels,
some recent works  on $tc$ FCNC \cite{peccei} and on
the effects of new heavy $Q=2/3$ isosinglets
on low energy physics \cite{barger,parada}.

We  assume the existence of  $N$ {\it new} $Q=2/3$ isosinglet
 $L$-handed quarks $U_L^o$, as can appear in
vector-like  multiplets $U_L^o$,  $U_R^o$
and that they are mixed with the  known $u$-type
quarks $u_L^o$,  $u_R^o$\footnote{
\baselineskip 16pt Our results
 hold  also when the $U_L^o$ isosinglets
appear in  mirror families $U_L^o$, $D_L^o$,  $(U^o  D^o)^T_R$.
However in this case the analysis is complicated
by the possible appearance of induced $R$-handed currents
leading to the new effective vertex $b_R s_R Z$.}.
The number $N$ of  $U_L^o$-$U_R^o$ pairs is irrelevant for our
 general analysis, and we will leave it unspecified.
$U_R^o$ and $u_R^o$, being both color triplet $Q=2/3$
isosinglet states, have the same gauge quantum numbers,
and then their couplings to the gauge bosons are not affected
by  mixing. This is not the case for the $L$-chirality states.
The vector $\Psi^{o}_{uL}=(u^o,U^o)_L^T$ of the doublet $(u^o)$ and
singlet $(U^o)$ gauge eigenstates
is related to the corresponding vector of the ``light" $(u)$
 and heavy $(U)$ mass eigenstates
$\Psi_{u L}=(u,\, U)_L^T$ through a unitary matrix ${\bf U}$
\begin{equation}
\pmatrix{u^o\cr U^o}_L = \ul
\pmatrix{u\cr U}, \qquad \qquad
\ul = \pmatrix{A &E\cr F & G}.
\label{mixing}
\end{equation}
Here $U=({U}_1,{U_2},\dots {U_N})^T,$ while  $u$ is the vector
containing the up, charm and top quarks.
The  unitarity of $\ul$ implies
\beq
A^\dagger A+F^\dagger F=A A^\dagger +E E^\dagger =I_{3\times 3},
\label{unitary}
\eeq
where $I_{3\times 3}= {\rm diag}(1,1,1)$.
We further introduce a unitary matrix $K$
for the $L$-handed $d$-type quarks
\beq
d^o_L = K d_L, \qquad  \qquad  K K^\dagger=K^\dagger K=I_{3\times 3}.
\label{Kd}
\eeq
After  introducing the $(3+N)\times 3$ matrix
\beq
P =\pmatrix{I_{3\times 3} \cr  0\cr},
\eeq
  the  Charged Current  (CC) coupled to the $W^\pm$
bosons can  be written as
\beq
{1\over 2}J^W_\mu = {\bar \Psi^o_{uL}}\,\gamma_\mu \,P \, d^o_L  =
{\bar \Psi_{uL}}\, \gamma_\mu \,\ul^\dagger P K \, d_L.
\label{CC}
\eeq
Then for the CC we can define the $(3+N)\times 3$  mixing matrix
\beq
V  = \ul^\dagger P   K = \pmatrix{V_u \cr V_U} =
 \pmatrix{A^\dagger K \cr E^\dagger K}.
\label{V}
\eeq
The $3\times 3$ Cabibbo Kobayashi Maskawa (CKM) matrix for the
light states $V_u = (A^\dagger K)$ is not unitary.
We note however that
 (\ref{unitary}) and (\ref{Kd})  imply
\beq
V^\dagger V=V_u^\dagger V_u+V_U^\dagger V_U =
 K^\dagger (A A^\dagger + E E^\dagger ) K = I_{3\times 3}.
\label{CCunit}
\eeq
In terms of the mass eigenstates,
the  Neutral Current (NC) coupled to the $Z$ boson reads
\beq
{1\over 2}J^\mu_Z =
{1\over 2}\bar\Psi_{uL}\, \gamma^\mu \ul^\dagger\,\Pt\,\ul\,\Psi_{uL}
-  s^2_W \, \bar\Psi_{u} \, \gamma^\mu Q \, \Psi_{u},
\label{NC}
\eeq
where  $s^2_W = \sin^2 \theta_W $ with $\theta_W$ the weak mixing
 angle, and
\beq
\Pt= P\times P^\dagger =  \pmatrix{I_{3\times 3} & 0 \cr 0 & 0\cr}
\label{Pt}
\eeq
is the projector on the $L$-handed $T_3=1/2$ isospin doublet states.
We note that in (\ref{NC})    the second term remains
flavor diagonal since the matrix of the electric charges
is proportional to the identity ($Q={2\over 3}\, I$).
In   contrast, for the isospin part of the current
the matrix of the isospin charges
${1\over 2} \Pt $  is not proportional to the identity, and
therefore the corresponding isospin  couplings
are FC. For the NC we can define
the following $(3+N)\times (3+N)$ mixing matrix
\beq
{\cal U} = \ul^\dagger \Pt \ul =
\pmatrix{A^\dagger A & A^\dagger E \cr E^\dagger A & E^\dagger E \cr}
\label{U}
\eeq
which is also not unitary. However from (\ref{Kd}) and (\ref{V}) and
 from the first equality in (\ref{Pt}), it follows that
\beq
{\cal U} = V\times V^\dagger.
\label{UVV}
\eeq
The matrix  of the FCNC couplings ${\cal U}$
satisfies  the following interesting properties:
\beq
{\cal U}{\cal U}^\dagger ={\cal U}^2={\cal U}\,; \qquad\qquad\qquad
 {\cal U}\, V =  V.                         \label{NCunit}
\eeq
The first equation tells us that the matrix ${\cal U}$
 is  idempotent.
This is not surprising, since from the first equality in
 (\ref{U}) it is clear that ${\cal U}$ can be straightforwardly
 interpreted as the projector operator on the $L$-doublets written in
 the basis of the mass eigenstates.
The second equation has a very important implication. Together with
(\ref{CCunit})
 it ensures that, in spite   of  the presence of
the FC couplings,  all the mass independent terms in the new penguin
diagrams, which carry the structure $V^\dagger\,{\cal U}\, V$,
cancel off.

The usual SM $L$- and $R$-handed chiral couplings of the $u$ quarks are
\beq
\el = {1\over 2} \, -  {2\over 3}\, s^2_W\, ,  \qquad \qquad\qquad
\er   =                   - {2\over 3} \, s^2_W.
\label{SM-couplings}
\eeq
However, from (\ref{NC}) we see that  the mixing
with the new isosinglets modifies the $L$-handed $u$ coupling,
and in particular introduces
 a FC term.  It is convenient to write the general
 $\Psi_{u_i}\Psi_{u_j}Z$ coupling  as
\beq
\elij = {1\over 2}\, {\cal U}_{ij}-{2\over 3}\, s^2_W \,\delta_{ij} =
\el\,\delta_{ij} \, + \, {1\over 2} \,({\cal U}_{ij}-\delta_{ij}),
\qquad\qquad i,j=u,c,t,1..N.
\label{couplings}
\eeq
In the second equation, the first term corresponds
 to a trivial extension of the SM  to $3+N$ $L$-handed doublets with
 no tree level  FCNC.
The second term accounts for the fact that the new
 $N$ states are isosinglets. The reason for writing the $L$-handed
 coupling as in (\ref{couplings}) is twofold. In the first place, we
 aim to compare the results for  the isosinglets case with  those  for
 a multi-generation model, for which only the first term is present.
Secondly,  through (\ref{couplings})
the derivation of the effective \bsz\ vertex in the presence
of the tree level FC couplings can be more  easily performed
 in two steps. The first step
traces trivially the  SM computation \cite{inami-lim} extended
to  $3+N$ generations. As a second step, we need just to compute
 the two  additional diagrams  depicted in Fig.1 which arise from the
second term in (\ref{couplings}).

The sum of the one loop penguin diagrams  which do not contain
any insertion of the FC couplings yields, in the Feynman gauge,
\beq
\Gamma^0_{\rm eff} =
{g^3\over (4 \pi)^2 c_W} \, \bar b_L \gamma_\mu s_L
\sum_j ( V^*_{jb}V_{js})\,\, \left[ X(x_j)+ Y(x_j)\right],
\label{GSM}
\eeq
where $c_W=\cos\theta_W$ and $x_j=m^2_j/M^2_W$ with $m_j$ the mass of
 the quark running inside the loop, and
\beq
 X(x_j) = - {5\over 4}\left[{1\over x_j-1} -
{ x^2_j \ln x_j  \over (x_j-1)^2} \right],  \qquad
Y(x_j) = {1\over 4}
\left[x_j - {2 x_j \ln{x_j} \over x_j-1}\right].
\label{X}
\eeq
This is the known  SM result as first given in \cite{inami-lim}.
The reason for introducing two different functions $X$ and $Y$ will
 become clear in the following.
 In (\ref{GSM}) the sum is taken over all the $Q=2/3$ quarks
which can appear in the loop. Then Eq. (\ref{CCunit}),  which is
 analogous to the unitarity of the CKM matrix in the SM and
implies  $\sum_j V^*_{jb}V_{js} =0$, ensures the correct
 cancellation of the same
set of mass-independent terms and divergences as in the SM case.
As is well known, the expressions in (\ref{GSM})
is not gauge-invariant by itself.
In order  to achieve a gauge-invariant result,  also the box diagram
 amplitudes  have to be taken into account.
For the two processes
$b\to s\, \bar l\, l$ (with $l = \nu$, $\ell^\pm$)
we are interested in,
the box diagram amplitudes have the  quark-mixing  structure
\beq
 {\cal M}^{\rm Box}_{\bar l l}\  \propto \
 \sum_j  (V^*_{jb}V_{js})\, W^l(x_j),
\label{Mbox}
\eeq
where in the Feynman gauge, and neglecting the masses of the
 charged leptons
\beq
 W^\nu(x_j) = 4\, W^{\ell^\pm}(x_j) =
  2 \left[{1\over x_j-1}
-{x_j \ln{x_j} \over (x_j-1)^2}\right].
\label{W}
\eeq
Now the sum
\beq
\sum_j  (V^*_{jb}V_{js})\,\left[
X(x_j)+ Y(x_j)+ W^l(x_j)\right],
\label{XYW}
\eeq
which appears in the full decay amplitude,  is a
physical, gauge invariant quantity.
 The experimental limits on  \bsll\ \cite{UA1} and   \bsnn\ \cite{yze}
 can then be used to set bounds on the masses and mixings of the
 possible new  $U$ doublets contributing to (\ref{XYW}).
However, other processes can be used to constrain
the same parameters. For example, via electromagnetic
penguins a quantity analogous to (\ref{XYW})
enters the expression for  the $b s  \gamma $ effective
 vertex \cite{inami-lim}, and thus the presence of the new doublets
affects also the rate for the radiative $b$ decay.

In the case we are analyzing here, the new
states are isosinglets, and  there are new contributions
from the FC couplings. The difference from the $N$ doublets case is
accounted for by the second term in (\ref{couplings}).
This term gives rise to the two additional diagrams depicted in Fig.1,
 which involve respectively loops of the $W$ gauge bosons and of the
 unphysical scalars~$\phi$.
At a first glance, both  the new  diagrams appear to be
logarithmically divergent.
However, the diagram involving the  unphysical scalars
$\phi$ is finite due to the presence of the $P_L$ chiral
projector (see Fig.1)
which reduces  the degree of divergence by a factor of 2.
After summing over
 all the $u$ and $U$ fermions, also the diagram involving the
$W$ loop is finite. In fact (\ref{CCunit}) and (\ref{NCunit}) imply
$\sum_{jk} V^*_{jb}\,\left({\cal U}_{jk}-\delta_{jk}\right)\,V_{ks}=0$
and thus all the terms independent of the $u$-quark masses (and in
particular the poles at D=4) cancel. Such a cancellation in the
presence of  this kind of tree level FC vertices
can be well regarded as a generalization of the SM GIM mechanism.
\begin{center}
\begin{picture}(10000,13000)
%\THINLINES
\THICKLINES
\put(5000,1000){\line(0,1){12500}}
\put(5000,1500){\vector(0,1){900}}
\put(5000,1500){\vector(0,1){4000}}
\put(5000,1500){\vector(0,1){8500}}
\put(5000,1500){\vector(0,1){11500}}
\put(5000,11500){\circle*{500}}
\put(5300,1500){$s$}
\put(5150,5800){$\Psi_{\! u_k}$}
\put(5150,8300){$\Psi_{\! u_j}$}
\put(5300,12200){$b$}
\put(-2000,8250){$({\cal U}_{jk}\,$-$\,\delta_{jk})\,P_L$}
\put(-6000,8200){$Z_\mu$}
\put(8300,7500){$W^-,~\phi^-$}
\put(5000,7500){\circle*{500}}
\drawline\photon[\W\REG](4900,7600)[1]
\drawline\photon[\W\FLIPPED](\photonbackx,7400)[1]
\drawline\photon[\W\REG](\photonbackx,7600)[1]
\drawline\photon[\W\FLIPPED](\photonbackx,7400)[1]
\drawline\photon[\W\REG](\photonbackx,7600)[1]
\drawline\photon[\W\FLIPPED](\photonbackx,7400)[1]
\drawline\photon[\W\REG](\photonbackx,7600)[1]
\drawline\photon[\W\FLIPPED](\photonbackx,7400)[1]
\drawline\photon[\W\REG](\photonbackx,7600)[1]
\drawline\photon[\W\FLIPPED](\photonbackx,7400)[1]
\drawline\photon[\W\REG](\photonbackx,7600)[1]
\drawline\photon[\W\FLIPPED](\photonbackx,7400)[1]
\put(3000,11500){$V^*_{jb}$}
\drawline\photon[\SE\REG](5050,11500)[1]
\drawline\photon[\SE\FLIPPED](\photonbackx,\photonbacky)[1]
\drawline\photon[\SE\REG](\photonbackx,\photonbacky)[1]
\drawline\photon[\SE\FLIPPED](\photonbackx,\photonbacky)[1]
\drawline\photon[\S\REG](\photonbackx,9050)[1]
\drawline\photon[\S\FLIPPED](\photonbackx,\photonbacky)[1]
\drawline\photon[\S\REG](\photonbackx,\photonbacky)[1]
\drawline\photon[\S\REG](\photonbackx,9025)[1]
\drawline\photon[\S\FLIPPED](\photonbackx,\photonbacky)[1]
\drawline\photon[\S\REG](\photonbackx,\photonbacky)[1]
\drawline\photon[\S\REG](\photonbackx,8975)[1]
\drawline\photon[\S\FLIPPED](\photonbackx,\photonbacky)[1]
\drawline\photon[\S\REG](\photonbackx,\photonbacky)[1]
\drawline\photon[\S\REG](\photonbackx,8950)[1]
\drawline\photon[\S\FLIPPED](\photonbackx,\photonbacky)[1]
\drawline\photon[\S\REG](\photonbackx,\photonbacky)[1]
\drawline\photon[\S\REG](\photonbackx,9000)[1]
\drawline\photon[\S\FLIPPED](\photonbackx,\photonbacky)[1]
\drawline\photon[\S\REG](\photonbackx,\photonbacky)[1]
\drawline\photon[\SW\FLIPPED](\photonbackx,\photonbacky)[1]
\drawline\photon[\SW\REG](\photonbackx,\photonbacky)[1]
\drawline\photon[\SW\FLIPPED](\photonbackx,\photonbacky)[1]
\drawline\photon[\SW\REG](\photonbackx,\photonbacky)[1]
\put(\photonbackx,\photonbacky){\circle*{500}}
\put(3000,\photonbacky){$V_{ks}$}
\end{picture}
\vbox{\hsize= 13.55truecm
\tenrm \baselineskip=12pt
% \quad
 \hskip -1.2 truecm FIG. 1.
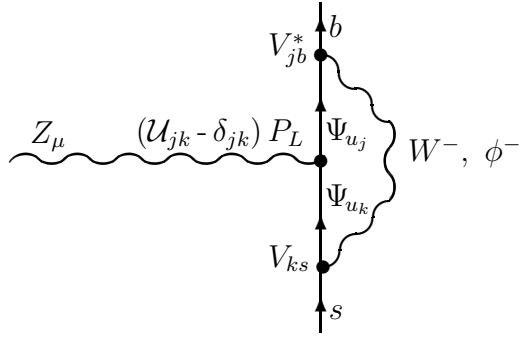
Electroweak penguin diagrams with  $W$ gauge bosons and  unphysical
 scalars $\phi$ which \break
\ \ \ include  the
flavor changing vertex \ ${\cal U}_{j\neq k}$.
The relevant mixing matrices appearing  at \break
 the vertices
are written explicitely, and $P_{L} = {1\over 2}(1 - \gamma_5)$ is the
$L$ chiral projector.\phantom{$ini$}
}
\end{center}
\medskip
The sum of the new $W$ and $\phi$ diagrams originating from the second
 term in (\ref{couplings}) which contains the  FC vertices reads
\beq
 \Gamma^{FC}_{\rm eff} =  \Gamma^{FC}_W +  \Gamma^{FC}_\phi =
{g^3\over (4 \pi)^2 c_W} \, \bar b_L \gamma_\mu s_L \,\sum_{j,k}\,
\left[V^*_{jb}({\cal U}_{jk}-\delta_{jk}) V_{ks}\right]\,\, Z(x_j,x_k),
\label{GFC}
\eeq
where
\beq
 Z(x_j,x_k) = {1 \over 4}{1\over x_j-x_k}
\left[{x_k-1\over x_j-1}x^2_j\ln x_j -
{x_j-1 \over x_k-1}x^2_k\ln x_k\right].
\label{Z}
\eeq
Since   no term proportional to
 $V^*_{jb}\,{\cal U}_{jk}\,V_{ks}$ can  arise from the box diagrams,
beyond being finite (\ref{GFC}) is also gauge invariant.
Apart from containing a FC part,  $ \Gamma^{FC}_{\rm eff}$ also
contains   flavor  diagonal terms proportional to $({\cal U}_{jj}- 1)$
 corresponding to  the limit of equal masses
\beq
\lim_{\ \ x_k\to x_j} Z(x_j,x_k)  =  Y(x_j)
\label{limit}
\eeq
and  $Y$ is given in (\ref{X}).
By combining (\ref{GSM})  and (\ref{GFC}) and by means of the
limit (\ref{limit}),  the effective \bsz\ vertex
 $\Gamma_{\rm eff}^{Z}=\Gamma^{0}_{\rm eff} +  \Gamma^{FC}_{\rm eff}$
can be recast as:
\beq
\Gamma_{\rm eff}^{Z} =
 {g^3\over (4 \pi)^2 c_W} \bar b_L \gamma_\mu s_L\,
\sum_{j}\left[
 (V^*_{jb}V_{js})\,  X(x_j) +
( V^*_{jb}\,{\cal U}_{jj}V_{js})\,  Y(x_j)  +
\sum_{k\neq j} (V^*_{jb}\,{\cal U}_{jk} V_{ks})\, Z(x_j,x_k)
\right].
\label{bsz}
\eeq
The replacement  $ X(x_j)\to  X(x_j) +  W^l(x_j)$ in (\ref{bsz})
 yields a physical quantity directly measurable in $b\to s \bar l l$
 decays. The first term inside the square brackets in (\ref{bsz}) is
 not affected by the fermion mixing.
The second term, which is also flavor diagonal,
 contains a quadratic $\sim x_j$ dependence
 which in the SM and in multi-doublet models represents
the  dominant contribution for very large masses.
 The mixing with the isosinglets reduces this contribution.
In fact,   being $\sum_{j=1}^N {\cal U}_{jj}=Tr (V^\dagger V)
=3$ and since the experimental bounds on the flavor diagonal $u_L$
 and $c_L$ mixings \cite{nrt} imply  ${\cal U}_{cc}\sim
{\cal U}_{uu}\sim 1$, we have
$\sum_{j=3}^N {\cal U}_{jj} \sim 1$. Hence
  the dependence on  the  large  masses
$m_t$, $m_{U_1}$\dots $m_{U_N}$
is weakened with respect to the doublet case
 ${\cal U}_{jj}=1$. Finally, the third term accounts for the
 additional effects of the FC  vertices.
Since for the $bs\gamma$  effective vertex
there are no diagrams analogous to the ones
depicted in Fig.1, the rate for \bsg\ is not sensitive to
the FC mixings.
Therefore, in the isosinglet case the indirect constraints
 on  $\Gamma^0_{\rm eff}$ from the experimental
measurement of $b\to s\gamma$ \cite{CLEO-bsg-in}
cannot be applied  to the full $bsZ$ effective vertex.

In the limit $|V_{tb}|\sim|V_{cs}|\sim 1$
no additional suppression beyond the loop  factor
can reduce the effect of  a large ${\cal U}_{tc}$, and then it is
interesting to study to what extent such
contribution can affect  the $bsZ$ effective vertex.
By means of (\ref{U})  the sum appearing  in
(\ref{GFC}), which accounts for the difference between the
isosinglet and isodoublet cases,  can be rewritten
as  a sum over $j\neq k$ terms involving only the ``FC function" $Z$
\beqa
&&\Gamma^{FC}_{\rm eff} \propto \sum_{j,k} V^*_{jb}\,
\left(\sum_{d}V_{jd}V^*_{kd}-\delta_{jk}\right)V_{ks}\, Z(x_j,x_k) =
  \cr
&& \sum_{j\neq k}\left[V^*_{jb}V_{js}
(|V_{kb}|^2+|V_{ks}|^2) +
 V^*_{jb}V_{jd}V^*_{kd}V_{ks}\right]\, Z(x_j,x_k).
\label{sumFC}
\eeqa
The second term inside the square brackets can  be neglected since
it always involves small intergenerational
mixings or a small value of $Z(x_u,x_c)$, resulting in
contributions never exceeding $10^{-3}$.
As for the first term, it has the standard
structure $V^*_{jb}V_{js}$ and  since no
ambiguities can arise from phase  differences,
it can be easily confronted
term by term  with the standard contributions in (\ref{GSM}).
Neglecting the possible  additional
suppression from
$(|V_{cb}|^2+|V_{cs}|^2)$, $(|V_{cb}|^2+|V_{cs}|^2)<1$
the maximum contribution from the FC $tcZ$ vertex reads
\beq
\left(
V^*_{tb}V_{ts}  + V^*_{cb}V_{cs}
\right) Z(x_t,x_c),
\label{tc}
\eeq
which is always smaller in absolute value than the
SM term, and  of opposite sign. For example, for
$m_t=180\,$GeV and $m_c=1.5\,$GeV we have   $Z(x_c,x_t)\simeq -0.51$,
while for the  leading  contribution to the corresponding term in
$\Gamma^0_{\rm eff}$
 we find $ X(x_t)+ Y(x_t)+ W^{\nu(\l^\pm)}(x_t) \simeq
2.59 \, (2.97)$.
Then the $t\,c$ FC  contribution to $\Gamma^Z_{\rm eff}$
interferes destructively with
$\Gamma^0_{\rm eff}$ thus {\it reducing} the $b\to s l l$
($l = \nu,\, \ell^\pm$) decay rates. Therefore
we can conclude that it is not possible to translate an
upper limit on these decays
into a bound on the strength of a $tcZ$ coupling induced by  mixing.

In the scenario we are analyzing here, beyond the effects of the $tcZ$
 coupling there are also other effects
related to the presence of the new heavy states, and it is worth
studying how these additional effects can influence the effective
 $bsZ$ vertex.

The  contributions to $\Gamma^0_{\rm eff}$
 of the heavy isosinglets yield the same  enhancement of the
effective vertex as for new isodoublets.
In fact, a very heavy isosinglet with sizeable couplings to
the $b$ quark would effectively play the role of a heavier
$t$ quark, thus enhancing $\Gamma^{0}_{\rm eff}$ and the overall
effective vertex. However,  we stress again that this situation would
 not  affect only the processes involving  $Z$ boson exchange
but also, and in a similar way, other processes like $b\to s \gamma$.
For the isosinglets, there are specific
additional effects from the FC contributions.
The effect of a FC mixing of the  heavy singlets $U$ with the  $c$
quark does not differ from the $t\,c$ case.  Since we always have
$Z(x_U,x_c)< 0$ and  $|Z(x_U,x_c)|< X(x_U)+ Y(x_U)+ W^{\nu(\l^\pm)}
(x_U)$ also these terms interfere destructively with the
corresponding terms in $\Gamma^0_{\rm eff}$, thus weakening the strength of
the effective $bsZ$ vertex  with respect to the doublet case.

If both $m_i$ and $m_j$ are $\gtrsim 150\, $GeV, then
the function $Z(x_j,x_k)$ is positive. Therefore
the contribution of a $U\,t$ coupling as well as of a
pair of new heavy states $U_1\,U_2$ adds constructively to
$\Gamma^0_{\rm eff}$.
However, in the limit $x_j >> x_k >> 1$ we have
$Z(x_j,x_k) \sim x_k \log x_j$ to be contrasted with the quadratic
enhancement $Y(x_j) \sim x_j $  appearing in  $\Gamma^{0}_{\rm eff}$.
Namely,  even for large masses, the new FC couplings
which are peculiar of the isosinglet case can induce only
small positive logarithmic deviations from the isodoublet case.
For example, for $m_U=500\,$GeV  and
$m_t=180\,$GeV the presence of a $\>{\cal U}_{Ut}$ term can enhance
the isosinglet case at most by 10 \% with respect to
the contributions of a  new doublet of equal mass.

In conclusion, we have shown that the presence  of a mixing induced
$tcZ$ vertex is expected to {\it lower} the rate for the decays
\bsll\ and \bsnn, and therefore it cannot be constrained by the
experimental limits on \bsll\ \cite{UA1} and \bsnn\ \cite{yze}.
More in general,  we have found that  the presence of $Q=2/3$
 isosinglet quarks cannot  yield any relevant enhancement of the
 $bsZ$ vertex with respect to the better known case of additional
 doublets, as from a fourth generation, which
to some extent is constrained by other rare processes as \bsg.
This suggests that
it is very unlikely that the peculiar effects  of a
mixing with isosinglets
will be observed in low energy processes as \bsll\  and
\bsnn\  with the precision in the foreseeable  future.

% \section{Acknowledgments}

\vspace{1cm}
\centerline{\large \bf Acknowledgments}
\vspace{0.4cm}
\noindent
It is a pleasure to thank Yuval Grossmann, Yossi Nir and Adam Schwimmer
for illuminating discussions and useful comments, and Ira Rothstein for a
 careful reading of the manuscript.

%%%%%%%%%%%%%%

%%%%%%%%%%%%%%%%%%%%%%%%%%%%%%%%%%%%%%%%%%%%
\end{document}